%% file: main.tex
\documentclass[conf]{new-aiaa}
\usepackage[utf8]{inputenc}

\usepackage{graphicx}
\usepackage{amsmath}
\usepackage[version=4]{mhchem}
\usepackage{siunitx}
\usepackage{longtable,tabularx}
\usepackage{svg}
\usepackage{tikz}
\usepackage{pgfplots}

\hypersetup{%
	colorlinks=true,
    linkcolor=blue,
    filecolor=blue,
    urlcolor=blue,
    citecolor=blue,    
    pdftitle={Development of a Structured Approach for Establishing Mission Engineering Requirements},
    pdfauthor={Taylor C. Fazzini and Daniel R. Herber},
}

\def\notesflag{0}

\newcommand{\xnote}[1]{\if\notesflag1
\textcolor{red}{\textit{#1}}
\else%
\fi}

\title{Development of a Structured Approach for Establishing Mission Engineering Requirements}

\author{Taylor C. Fazzini\footnote{Doctoral Student, Systems Engineering, and AIAA Senior Member. \href{mailto:Taylor.Fazzini@colostate.edu}{Taylor.Fazzini@colostate.edu}} and Daniel R. Herber\footnote{Associate Professor, Systems Engineering, and AIAA Senior Member. \href{mailto:Daniel.Herber@colostate.edu}{Daniel.Herber@colostate.edu}} }
\affil{Colorado State University, Fort Collins, CO, 80523}

\begin{document}

\maketitle

\begin{abstract}
This paper addresses the question: How can mission effectiveness be systematically defined or approximated in the absence of customer requirements? Legacy requirements engineering frameworks presuppose customer input to define specifications but leave a gap in the process when stakeholder input is ill-defined or missing. Rapid build and development programs (such as military acquisition, space assets, infrastructure projects, etc.) often see requirement and objective evolutions throughout the proposal process, so a more adaptive method is needed. To address this gap, a structured approach is proposed that decomposes mission intent into mission context, functions, constraints, critical dimensions, effectiveness attributes, and architecture alternatives. This method conducts a mission feasibility assessment, prioritizes mission-critical dimensions using Best-Worst Scaling, and introduces a mission complexity factor to quantitatively understand the impacts of external mission difficulties, technology maturity, evidence and confidence standards, and mission utility. The resulting method provides a traceable basis for deriving Tier 1 and 2 requirements. The approach is structured to support future Unified Architecture Framework (UAF) and Systems Modeling Language (SysML) artifact integration. The proposed framework is demonstrated using a notional close air support mission example.
\end{abstract}

\section*{Key Nomenclature}
{\renewcommand\arraystretch{1.0}
\noindent\begin{longtable*}{@{}l @{\quad=\quad} l@{\qquad} l @{\quad=\quad} l@{}}
DE & Digital Engineering & MBSE & Model-Based Systems Engineering \\
BIBD & Balanced Incomplete Block Design & MCF & Mission Complexity Factor \\
BWS & Best-Worst Scaling & ME & Mission Engineering \\
CAS & Close Air Support & MFA & Mission Feasibility Assessment \\
DoD  & Department of Defense & SME & Subject Matter Expert \\
DoDAF & DoD Architecture Framework & SysML & Systems Modeling Language \\
EBD & Effectiveness-Based Design & TC & Technology Complexity \\
KPP & Key Performance Parameter & TRL & Technology Readiness Level \\
MAUT & Multi-Attribute Utility Theory & UAF & Unified Architecture Framework \\
\end{longtable*}}

\section{Introduction}
\lettrine{I}{n} the earliest phases of aircraft development, engineers must make decisions that have far-reaching, and often not well-understood, impacts on the system's design. This lack of clarity brings significant technical risk, such as evaluation of alternatives made with invalid assumptions and trade spaces artificially limited due to a misunderstanding of customer desire. Maintaining small teams on proposal efforts often leads to experts in niche disciplines becoming involved late in the process, driving both schedule and cost changes when impacts and risks are finally realized. Systems engineering has become a major driver in understanding downstream impacts by pulling integration planning and requirements management forward in the design process. 

\xnote{This paragraph: establishes the military flavor of the paper w.r.t digital engineering}
While systems engineering has existed in military applications since the 1940s to aid in the development of systems analysis, missile design, and communication systems \cite{goode_system_1957, schlager_systems_1956}, the subdiscipline of digital engineering (DE) is relatively nascent. Rapid technical accelerations in the field have led to sub-specialties within DE, including Digital Threads \cite{verma_transforming_2023}, Digital Twins \cite{chelliah_digital_2020}, and Agile Systems Engineering Methodologies \cite{granados_digital_2023}. The speed of development into the field of DE has resulted in the United States Department of Defense (DoD) issuing guidance and revisions each time a new branch of systems engineering is defined \cite{noauthor_dod_2024, noauthor_engineering_2020, noauthor_digital_2023,noauthor_requirements_2022}. This piecemeal approach is challenging for organizations who are financially committed to their current suite of tools and especially for those already in the middle of a design cycle. The US Defense Acquisition University (DAU) provides context into the complexities of DE that have made adoption challenging for DoD programs with heavily classified and siloed information. They define DE as ``an integrated digital approach that uses authoritative sources of systems' data and models as a continuum across disciplines to support lifecycle activities from concept through disposal'' \cite{noauthor_dau_nodate}. 

\xnote{This paragraph: sets up the integration of digital engineering into air force}
Although DE, as a sub-discipline of systems engineering which uses digital tools and data to support the full lifecycle of a system, was not a ground-breaking topic in 2020, it was not until then that the Department of Defense first released guidance requiring the use of comprehensive engineering plans for all major acquisition programs, including systems engineering plans, mission engineering, and technical risk assessments \cite{noauthor_engineering_2020}. In 2023, they expanded this requirement to include DE practices for defense acquisition programs \cite{noauthor_digital_2023}. A paradigm shift is needed to maintain agility in the acquisition process while adapting to DE mandates \cite{noauthor_digital_2019}. This shift is especially difficult in an age of increasingly complex systems, where all disciplines (including design, test, deployment, and sustainment) must be digitally traced and constantly updated in a model-centric environment. Roper provides guidance on the transition specifically related to the US Air Force's acquisition strategy for new programs \cite{roper_take_2020, roper_bending_2021}. While the DoD mandates apply to all service branches and types of new products, this research will primarily focus on the processes followed for aircraft development programs in the US Air Force and Navy. Overlap into the development of other defense products (such as ships, satellites, ground vehicles, etc.) is likely but is outside the current scope of research. 

\subsection*{Mission Engineering}

\xnote{This paragraph: defines mission engineering}
Mission engineering (ME) is a systems engineering discipline most often leveraged by military programs, sitting at the intersection of operations analysis and systems of systems design. The DoD's Mission Engineering Guide (MEG) describes it as a cross-disciplinary process focusing on the design and analysis of the entire mission objective rather than its constituent platforms or capabilities and often includes considerations for operational execution, logistics, environmental context, and full force portfolio \cite{noauthor_mission_2023}.

\begin{figure}[tb]
    \centering
    \includegraphics[scale=0.5,trim={1.5cm 6cm 1.5cm 0},clip]{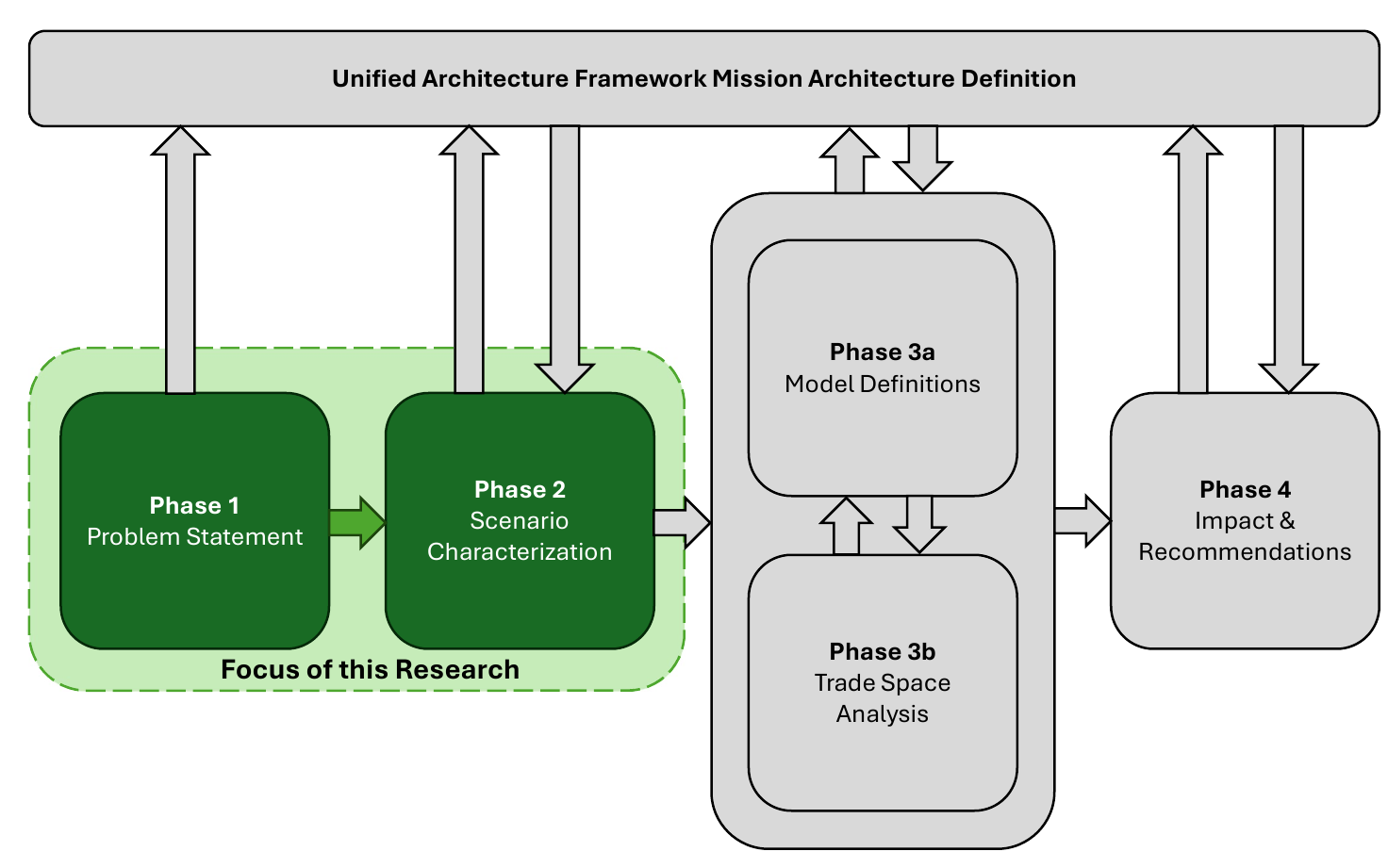}
    \caption{Mission Engineering Process with UAF Integration and Focus of This Research}
    \label{fig:me_process}
\end{figure}

At its core, ME is concerned with how an asset interacts with its environment and the outside world rather than just the design and manufacturing of the asset itself. ME enables analysts to consider the wider-ranging impacts a new platform can have on an entire conflict or objective, encouraging campaign-level trade space exploration, early identification of capability gaps, and better understanding of platform integration. This operational context is often used to provide decision-making support to individual disciplines, as a primary goal of mission engineering analysis is to determine the right combination of technologies, processes, and systems that meet the intended mission objectives. By adding technical rigor to a classically brainstorming and intuition-focused discipline, ME provides a structured framework to quantitatively analyze mission-level design to ensure platforms meet the operational needs of a contested environment \cite{hause_mission_2024}. This mission-centric perspective links the technical solutions of a holistic system directly to strategic and operational outcomes by providing the necessary mission evaluation context. Effectiveness-based design (EBD) leverages ME digital engineering processes into the design phase of a program to shift primary program metrics from classical aircraft performance figures of merit to mission effectiveness metrics \cite{harper_effectiveness-based_2021}. A mission engineering process integrated with digital engineering is shown in Figure~\ref{fig:me_process}. 

\subsection*{Requirements Engineering}

\xnote{This paragraph: defines requirements engineering}
Requirements engineering (RE) is another systems engineering specialty that develops and manages all requirements for a system \cite{noauthor_nasa_nodate}. It ensures that all customer needs and constraints are clearly addressed in the project through traceable and actionable requirements. A new project's requirements should not only describe what a system needs to accomplish, but also how it should achieve its objectives. It can describe technical, logistic, business, legal, or any other factors necessary to complete the mission. When done well, requirements engineering helps to maintain clear lines of communication between all stakeholders and identifies challenges and opportunities early in the project.
Historically, requirements engineering frameworks relied heavily on customer inputs to derive system specifications. In an ideal world, all requirements are supplied by the customer and fed directly into a requirements management tool. Much of the DoD uses a hierarchical Top-Level Requirements (TLRs) system to define three requirement tiers prior to the release of a Capability Development Document (CDD) \cite{lescher_top_2021}: 
\begin{itemize}
    \item Tier 1 - define the core of a program and often become the Key Performance Parameters (KPPs)
    \item Tier 2 - lower level of importance and higher flexibility; often become the Key System Attributes (KSAs) 
    \item Tier 3 - lowest importance and highest flexibility; often become additional performance variables
\end{itemize}
For modern programs, suppliers often try to ``shape'' requirements prior to a contract award to favor their solution over competitors. Realistically, requirements and their priority ranking evolve continuously throughout the early stage of program conception and the engineering development process. As systems become increasingly more complex and interconnected, the number of requirements on a program balloons into a problem that requires specific Requirement Management Tools (RMT) to maintain traceability \cite{wheaton_digital_2024}.

\subsection*{Model-Based Systems Engineering}

\xnote{This paragraph: defines the DoD-centric UAF}
Model-Based Systems Engineering (MBSE) is defined by the DAU as ``formalized application of modeling to support system requirements, design, analysis, verification, and validation activities beginning in the conceptual design phase and continuing throughout development and later system life-cycle phases'' \cite{noauthor_dau_nodate}. It is a method of DE that utilizes models to represent both physical and functional elements of a system, rather than traditional document-based approaches. MBSE uses tools to manage requirements from the systems-of-systems (SoS) level down to the individual component level to maintain full traceability within a system \cite{madni_modelbased_2018}. The DoD Architecture Framework (DoDAF) was a standardized MBSE framework first released by the DoD in 2003 to support decision-makers by providing both detailed architecture designs and high-level overviews \cite{department_of_defense_chief_information_office_dod_nodate}.

The DoDAF has since been succeeded by the Unified Architecture Framework (UAF), which was first released in 2017 and most recently updated to Version 1.2 in 2022. The UAF incorporates input from the DoDAF, the United Kingdom's Ministry of Defence's Architecture Framework (MoDAF), and the North Atlantic Treaty Organization's (NATO) Architecture Framework (NAF) \cite{uaf_v1.3}. Pulling from its framework pedigree, the UAF utilizes ten viewpoints, or stakeholder domains: architecture management, strategic, operational, services, personnel, resources, security, projects, standards, and actual resources. These form the vertical axis of the UAF grid system. The horizontal axis is composed of eleven aspects, or characteristics: motivation, taxonomy, structure, connectivity, processes, states, sequences, information, constraints, roadmap, and traceability. The intersection of each of these viewpoints and aspects forms a view specification. Seventy-one view specifications exist in the UAF, so gaps are present in the grid. It is the responsibility of the architects to determine which specifications fit the purpose of the program, as not all models may add value to the system. The current version of the UAF, v1.3, adds additional mission engineering capability to the framework \cite{uaf_v1.3}. 

\xnote{This paragraph: introduces SysML}
The Systems Modeling Language (SysML) is a standardized modeling language by the Object Management Group \cite{object_management_group_systems_nodate}.  SysML contains nine different types of diagrams, many of which have legacy roots from the Unified Modeling Language (UML): activity, block definition, internal block, package, parametric, requirement, sequence, state machine, and use case \cite{finance_sysml_2010}. It provides the ability to model system requirements, behaviors, and solution structure with a flexible architecture that is easily modifiable by system architects to best meet their needs. SysML helps to shift away from the legacy process of document-centric systems engineering practices towards a model-centric engineering method that improves consistency and traceability between disciplines, system components, and disparate teams by maintaining a universal baseline of the project solution \cite{kausch_enhancing_2025}. While it is not tailor-built as an RMT, recent work has shown that it is a suitable method for Model-Based Structured Requirements (MBSR) when meta-models and customization are included \cite{herber_model-based_2022, wheaton_digital_2024}. The most recent specification revision of SysML, Version 2, was released in September 2025 \cite{object_management_group_about_nodate}.

\subsection*{Overview}

\xnote{This paragraph: summarizes the scope of this paper}
Although adoption of modern system engineering methods is becoming more widespread, especially in the areas of software development and systems integration, best practices for their use in the areas of conceptual aircraft design and mission engineering are sparse. The permeation of digital engineering into the aerospace and defense industry over the past decade has highlighted deficiencies in the traditional design cycle, authoritativeness of models and data, and -- at the crux of solid systems engineering doctrine -- requirements traceability. While the MEG's process gives a general overview of a mission analysis architecture, it makes the assumption that customer requirements are already understood. There is currently a lack of a codified method of defining a mission and its corresponding requirements in lieu of deep customer understanding or previously established specifications. For this research, UAF is used as the architecture framework and domain structure for organizing mission context, operational performers, activities, exchanges, constraints, and traceability. SysML provides the underlying modeling language for parametrics, structure, behavior, and requirements, while UAF provides mission viewpoints used to organize those constructs. The proposed method treats each step as a model transformation from mission-level context to model-based requirements. While full implementation of the UAF architecture artifacts represented through SysML model elements is reserved for future work, this paper identifies the artifacts needed to develop context and formalize relationships needed for early conceptual design. This paper explores the decomposition of mission objectives into mission effectiveness parameters, the development of mission-relevant requirements, and demonstrates a notional use case for a close air support (CAS) mission. 

\begin{figure}[tb]
    \centering
    \includegraphics[scale=0.45,trim={0 7.5cm 0 6.5cm},clip]{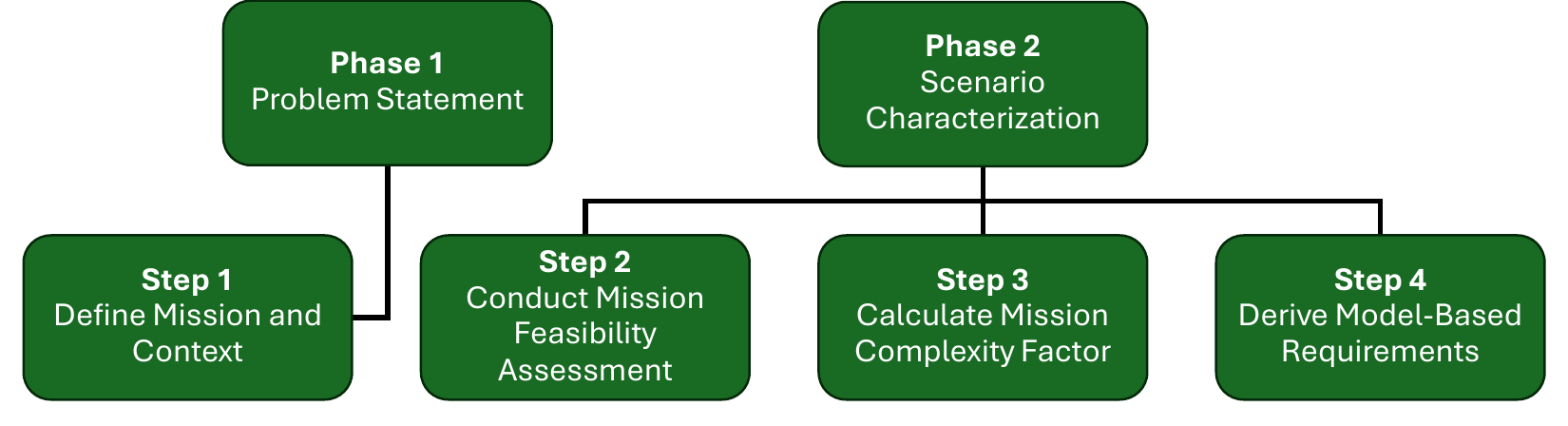}
    \caption{Relationship of ME Process to New Approach Flow}
    \label{fig:activity}
\end{figure}

\section{Mission Definition (Step 1)}
Mission engineering of a new program often begins with ambiguous, incomplete, or non-existent requirements from the customer. Due to the MEG's lack of requirement definition processes, a new framework is needed to bridge the gap between requirements and mission engineering. Figure \ref{fig:activity} shows the approach developed by this research, which codifies Phase 1 and Phase 2 of the ME process. Step 1 of this structured approach is to establish the missions that a solution needs to complete. While not formally documented, in the current military acquisitions environment, contractors are often expected to provide a solution without receiving written requirements from the customer. Without an existing working relationship with the customer, it is difficult to predict future mission and technology gaps that need to be filled. Customers may solicit contractors for insight into discriminators, but this can skew future competitions in the direction of the contractor that provided the insight. To combat the continuous cycle of offering new capabilities to a customer without knowing if they are relevant to their mission environments and future campaigns, this section will establish a methodical process for defining a mission, defining the effectiveness characteristics contributing to the mission, and proposing a new method of evaluating the feasibility and complexity of defined missions.

\subsection*{Step 1: Defining Mission Context}
\label{sec:mission-context}

\xnote{This paragraph: introduces the definition of a mission and associated terms}
The DoD's MEG defines a mission as a ``task, together with the purpose, that clearly indicates the action to be taken and the reasoning behind it'' \cite{noauthor_mission_2023}. For the purpose of this research, the definition of mission will follow the MEG's guidance -- it is a singular task to be accomplished. Alongside it is the mission context, which will answer the `who', `what', `when', `where', and `why' that help to fully understand the purpose and background of the mission. In this new, structured approach to developing requirements, these questions must be answered before the `how' is considered. Step 1 of this new approach is not complete until the context of the mission is defined, as the problem cannot be clearly understood without also understanding the environment surrounding it. For this research, the mission thread is the component that defines and documents the mission context. Similar to the `flight profile' in most other aerospace disciplines, the mission thread is the series of steps and events that must occur to successfully complete a mission \cite{noauthor_mission_2023}. This research assumes the mission thread can include both system and human actions, including flight profile, human input commands, autonomous system actions, crew actions, decision points, information exchanges, environmental triggers, and end states.

For example, a mission could be to ``take an aerial picture of a wildfire boundary'' which has many sub-tasks required to successfully accomplish that task. Components of the mission thread may include planning the flight route to avoid patches of heavy smoke and low visibility; the optimal direction to arrive on station to minimize interference with aerial firefighting routes; geolocating the boundary of the wildfire; a sensor operator calibrating and collecting images of the wildfire; contingencies in case of unsafe winds, smoke, or competing air traffic; and the total amount of fuel needed to safely arrive back at the landing airfield. While the big picture of the mission can be easily understood by decision-makers, the mission thread helps analysts, operators, and engineers ensure that the mission is actually achievable by breaking everything into sub-components. 

\xnote{This paragraph: introduces ov-1's and visual mission engineering depictions.}
Another way of communicating with decision-makers and stakeholders about the mission and its mission thread is using a visual depiction of the tasks at hand. An OV-1 is a top-level operational view in the UAF that visually shows how all elements of a mission interact to support the common goal. It traditionally includes communication linkages and interactions between architectures, the environment, and external entities \cite{department_of_defense_chief_information_office_dod_nodate}. Figure \ref{fig:ov1} below shows a nominal OV-1 for a close air support mission against a ground-based threat. Adversary ground vehicles and troops are shown en route to friendly ground troops (blue). Friendly air, space, and ground assets are shown with their communication links (yellow), control links (white), and tracking cones to adversary (red). OV-1 visuals help to show the interconnectedness of a potential systems-of-systems solution to easily communicate mission objectives with all relevant stakeholders. Toward the beginning of a conceptual design phase, an OV-1 can be used to show a generalized version of the mission solution with all its technical capabilities and benefits before the details of the design are fully developed. 

\begin{figure}[tb]
    \centering
    \includegraphics[scale=0.485]{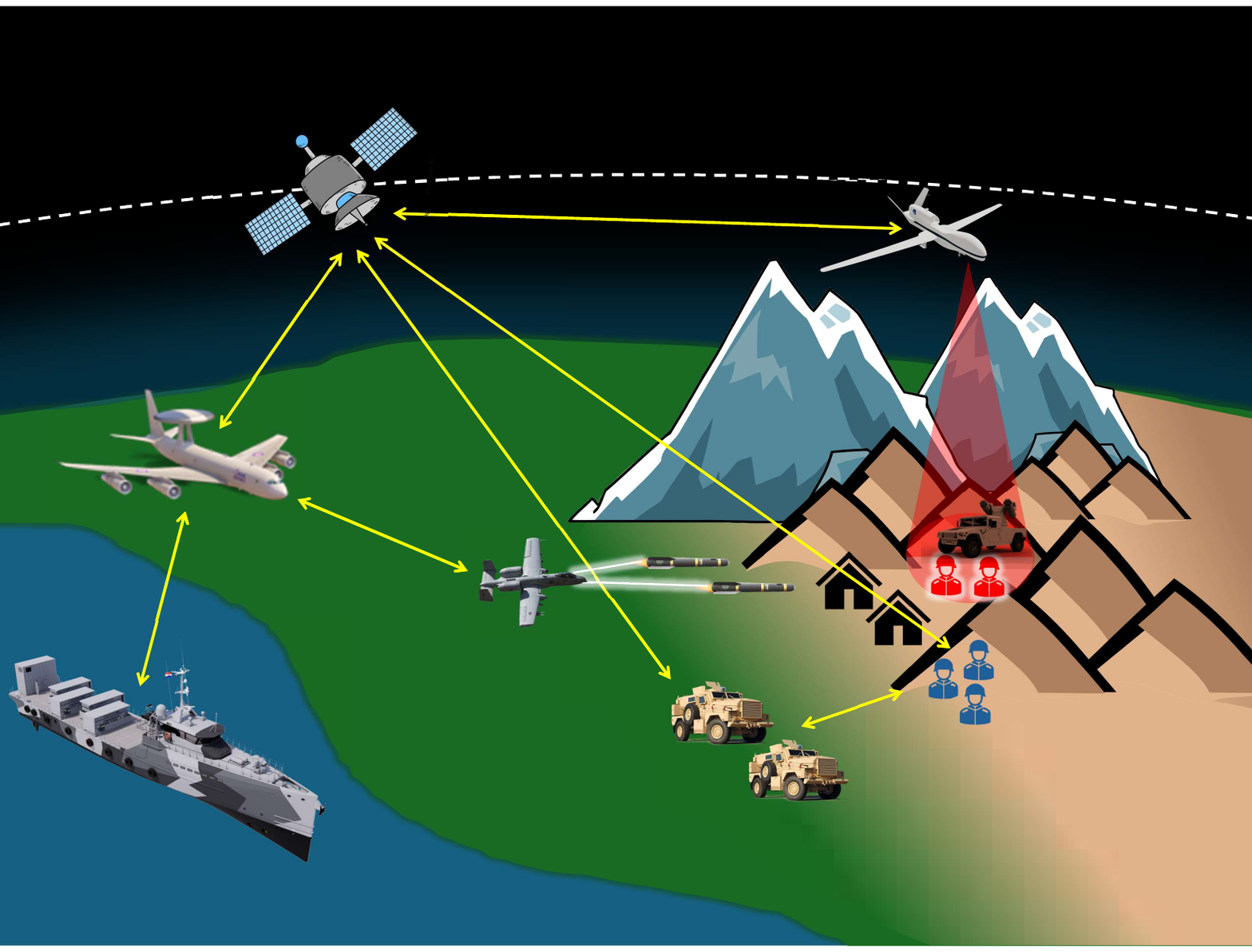}
    \caption{Example of Mission-Level OV-1 for a Close Air Support Mission}
    \label{fig:ov1}
\end{figure}

In the proposed method, the OV-1 and mission thread serve as the initial UAF architecture artifacts that can be represented through SysML model elements, linking mission objectives, information pathways, performers, environmentals, and end states.

\section{Mission Feasibility Assessment (Step 2)}
Once all relevant missions for a system are clearly documented and understood, Step 2 - the brainstorming phase of `how' - can be considered from Figure~\ref{fig:activity}. Individual technologies do not yet need to be defined at this stage; rather, the intent is to characterize a mission's feasibility envelope as a boundary of what is achievable given the current state-of-the-art, operational constraints, and known capabilities. First, the driving factors for mission success must be identified. When considering the effectiveness of a mission, ME often uses a grouping of words commonly referred to as ``-ilities''. The Defense Acquisition University defines these as `the operational and support requirements a program must address' \cite{noauthor_dau_nodate}. To summarize, they are a collection of quality attributes that help to describe how effectively a solution performs against its expected objectives. They often describe nebulous, operationally relevant concepts and uncertainties rather than traditional design metrics (such as weight, speed, length, etc.). For ease of reference, ``-ilities'' will be referred to as critical dimensions for the remainder of this work.

The integration of the critical dimensions into a solution's design process helps engineers assess the real-world performance of a system and ensure that design decisions are appropriately influenced by the long-term operational implications, costs, and flexibility of the system. This forward inclusion helps establish evaluation criteria for ME analysis, system resiliency, and traceability back to mission objectives. A selection of the most commonly used defense critical dimensions is shown in Table~\ref{tab:ilities}. 

\begin{table}[tb]
\begin{center}
\caption{Definitions of Common Mission-Centered ``-ilities'' \cite{noauthor_dau_nodate}}
    \begin{tabular}{|>{\raggedright}p{2.2cm}|m{12.5cm}|} 
  \hline
  \textbf{Critical Dimensions} & \textbf{Definition from Defense Acquisition University}  \\
  \hline
  Affordability & determination that the Life Cycle Cost (LCC) of an acquisition program is in consonance with the long-range investment and force structure plans of the DoD or individual DoD components \\ 
  \hline
  Availability & measure of the degree to which an item is in an operable state and can be committed at the start of a mission when the mission is called for at an unknown (random) point in time. Three categories:  Inherent Availability (Ai), Achieved Availability (Aa), and Operational Availability (Ao) \\ 
  \hline
  Interoperability & ability of systems, units, or forces to provide and accept services from other systems, units, or forces, and to use the services so exchanged to enable them to operate effectively together \\ 
  \hline
  Lethality & probability that a weapon will destroy or neutralize a target \\ 
  \hline
  Maintainability & ability of an item to be retained in, or restored to, a specified condition when maintenance is performed by personnel having specified skill levels, using prescribed procedures and resources, at each prescribed level of maintenance and repair \\ 
  \hline
  Modularity & design where functionality is partitioned into discrete, cohesive, and self-contained units with well-defined interfaces that permit substitution of such units with similar components or products from alternate sources with minimum impact \\ 
  \hline
  Survivability & capability of a system or its crew to avoid or withstand a manmade hostile environment without suffering an abortive impairment of its ability to accomplish its designated mission \\ 
  \hline
  Susceptibility & degree to which a device, equipment, or weapon system is open to effective attack as a result of one or more inherent weaknesses. Susceptibility is a function of operational tactics, countermeasures, probability of enemy fielding a threat, etc \\
  \hline
  Utility & state or quality of being useful militarily or operationally. Designed for or possessing a number of useful or practical purposes rather than a single, specialized purpose \\
  \hline
  Vulnerability & characteristics of a system that cause it to suffer a definite degradation (loss or reduction of capability to perform the designated mission) as a result of having been subjected to a certain (defined) level of effects in an unnatural (man-made) hostile environment \\
  \hline
    \end{tabular}
    \label{tab:ilities}
\end{center}
\end{table}

For a mission whose objective is to provide CAS to ground troops, survivability, lethality, and reliability are likely to be high-ranked in importance to successfully complete the mission. Because there will be many different solutions that can all accomplish the singular task of the mission, multiple options should be developed to conduct an operational trade study. 

\subsection*{Step 2.1: Narratively Define Success}
In lieu of quantitative success metrics, an operational narrative is used to create a qualitative success test that can be consistently judged. For the CAS mission, a success statement could be ``Support is successful if it arrives within the ground commander's decision window, provides the intended munitions and effects, and does not create unacceptable civilian or ground troop risk''. While this does not quantify the time frame of the commander's window, the amount of effects, or what qualifies as unacceptable risk, it provides a statement that can be judged for pass/fail. Alongside the success statement, additional definitions should provide more context to evaluate success. For the CAS mission, these may include ``target validation meets established confidence standards'', ``mission abort remains available throughout the duration of the attack window'', or ``deconfliction with other assets can be maintained at nominal complexity levels''. Parallel with the success statement, these definitions help to clearly characterize what makes a mission successful without the need for specific quantitative criteria. This step is complete when the success statement and supporting definitions have been translated into mission success criteria that can be modeled as effectiveness measures, constraints, or requirement elements.

\subsection*{Step 2.2: Define Platform-Agnostic Mission Functions and Constraints}
Using the mission thread developed in Section~\ref{sec:mission-context}, a preliminary list of constraints is documented that can be ranked by difficulty. First, all major mission functions should be established to more easily understand which constraints and critical dimensions may be relevant to the mission. In general, these should be verbs that describe an action the mission must complete to be successful without implying a specific technology or platform. For the CAS mission, potential mission functions include \textbf{access} (enter contested mission area), \textbf{coordinate} (with both people and systems), \textbf{sense} (identify, classify, or interpret information on adversary and ally ground troops), \textbf{decide} (on tasking, prioritization, and timing), \textbf{deliver} (payloads or effects), or \textbf{recover} (extract and return to base). 
These functions should be captured as platform-agnostic operational activities with a UAF mission architecture and can represented with SysML activity constructs, along with one to two sentences describing each function. 

Next, categories of mission constraints are defined that could impede each mission function, along with a three-tiered ranking of constraint level as low/medium/high. Although these constraints can vary by mission environment, they will be used to identify potential pitfalls while not to develop a singular solution. From the constraints, a mission envelope table can be created. For the CAS mission functions, examples include: 
\begin{itemize}
    \item Sense - environment severity: benign/harsh/extreme
    \item Decide - information quality: known/ambiguous/deceptive
    \item Coordinate - human dependency: autonomous/human intervention/human controlled
    \item Recover - failure tolerance: low/medium/high
\end{itemize}
This step is complete when the mission functions and constraints are captured in a mission function model and a mission envelope table.

\subsection*{Step 2.3: Rate Mission Inherent Difficulty}
With the knowledge base of the mission thread developed, the external factors of the mission are considered to understand the \textit{overall difficulty of the mission}. For each of the rankings described below, a scale of 1-5 is used, with 1 being the `least' and 5 being the `most' of the relevant assessment. First, using the mission envelope table developed in the previous step, the mission engineer chooses the level that seems most likely for the mission using the low/medium/high scale established. For this analysis, the levels are transformed into numeric parameters: low is 1, medium is 3, and high is 5. If every constraint is listed as ``high'', it is an early indication of an inherently difficult mission.

Next, the \textit{operational complexity} is ranked from 1 to 5. Factors to consider for operational complexity in the CAS mission could include airspace control, necessary support units, handoffs, or dynamic environment changes. 
Following this, the \textit{uncertainty and ambiguity} of the mission are scored using the same low-to-high scale as previously defined. Uncertainty and ambiguity rankings should consider whether they include any pieces of information that are unknown or change rapidly. 

The final factor considered in mission inherent difficulty is the \textit{consequence of failure}. In high-stakes missions, there are often catastrophic effects of failure. These can range from operational impacts, like the unintended loss of human life, to political impacts, such as causing a new conflict between two countries. 

In a model-based implementation, these scored factors are represented as SysML value properties and connected through a  parametric constraint model to calculate the mission inherent difficulty, $D$, as the normalized summation of all factors.

\subsection*{Step 2.4: Rank Critical Dimensions}
In this step, relevant critical dimensions are identified and ranked by importance to the mission. First, critical dimensions are identified. For the CAS example, critical dimensions could include reliability, lethality, survivability, maneuverability, and modularity. As less important dimensions will be identified through the ranking process, the analyst should list as many as could be relevant when choosing critical dimensions for evaluation. 

One ranking procedure is the Maximum Difference Scaling (MaxDiff) survey method, which is a type of Best-Worst Scaling (BWS) that provides an information-rich discriminatory ranking system with uncertainty quantification \cite{maxdiff_ch5}. Research has found that BWS methods outperform Likert scales when discrimination between attributes is the primary priority rather than group differences \cite{bws_methods}. 
MaxDiff methods are broken into three cases, which classify problem sets by whether the items being ranked are objects, profiles, or multi-profile cases. Each case has corresponding governing equations. For this work, Case 1, which classifies the items as factors, is most appropriate to produce a full ranking of critical dimensions with relative importance scores. For each task, the respondent is given a different combination of the factor set; ideally, each factor appears an equal number of times. The probability $P$ of choosing factor $i$ as the most important option and factor $i'$ as the least in a set $X$ is given by: 
\begin{align}
    P_X(i,i') = \frac{\mathrm{exp}(v_i-v_{i'})}{\displaystyle\sum_{j\in X}\sum_{k\in X,k\neq j} \mathrm{exp}(v_j-v_k)}
\end{align}

\noindent where $v_i$ and $v_{i'}$ are the latent utility values for the selected most-important and least-important critical dimensions, $X$ is the set of critical dimensions shown in the task, and $j$ and $k$ index all ordered best-worst pairs in $X$.

For a total of $N_c$ critical dimensions selected to be analyzed, each task of the survey includes the same number, $S$, of critical dimensions. The number of times a dimension is shown is represented by $r$. Following best practices discussed by Chrzan and Peitz \cite{chrzan}, each critical dimension should appear at least 3 times, and 4 dimensions should be shown for each task. A larger number of appearances increases precision, so $r$ should be increased until a total task limit, $N_{tasks}$, is reached. The task limit is given by: 
\begin{align}
    N_{tasks} = \frac{N_c*r}{S}
\end{align}

The most common method to ensure each critical dimension occurs an equal number of times is to use a Balanced Incomplete Block Design (BIBD) \cite{louviere_2013}. Unlike fractional factorials, BIBDs ensure a pair of choices occurs equally often as well. Although specific BIBDs also exist to provide order parity, studies have been inconclusive on the difference in outcomes when order is strictly controlled. As the number of factors increases, the task load for the respondent can become too large. To maintain a reasonable number of tasks, if the quantity of the total number of tasks exceeds 20, critical dimensions should be split into domains to lessen the total workload of the survey respondent. 

Once this structured survey for critical dimensions is created, a cohort of 3-7 subject matter experts (SMEs) independently evaluate each task by choosing the most and least important critical dimensions with respect to the success of the mission as defined in Step 1 of the MFA process above. The SMEs will provide further justification for their judgment in later steps. 

The MaxDiff responses are analyzed using a Hierarchical Bayes multinomial logit model. For this, each SME's responses for the tasks are estimated as a vector of latent utilities of critical dimensions. At the SME level, the probability of selecting one dimension as most important and another as least important is modeled as the difference between their utilities; at the group level, the vectors are assumed to come from a multivariate normal distribution. This allows for the vectors to be informed by other experts while also preserving their individual information. Posterior means are then used to derive the mission-level utility weights $w_i$ for each critical dimension using the following:
\begin{align} \label{eq:wi}
    w_i = \frac{\mathrm{exp}({v_i})}{\displaystyle \sum \mathrm{exp}({v_j})}
\end{align}

This step is complete when each critical dimension is rank-ordered and represented as a weighted mission effectiveness attribute for use in the mission utility model.

\section{Mission Complexity Factor (Step 3)}
In Step 3 from Figure~\ref{fig:activity}, a mission complexity factor will be defined and calculated that incorporates both the previous mission feasibility assessment and a quantitative evaluation of the mission utility impact of driving technologies needed to accomplish the task. The mission complexity factor will be a single value that ranks each mission for a new program by the difficulty of the mission and the potential of technology infusions to alleviate said difficulty.

\subsection*{Confidence Levels}
The same confidence levels used by the intelligence community are used here to convey analytic assessment of each ranking in the previous step \cite{intel_guide}:
\begin{itemize}
    \item High confidence: based on high-quality information, supports solid judgment
    \item Moderate confidence: information is plausible but not sufficiently corroborated to justify high confidence
    \item Low confidence: analyst has significant concerns with sources, or information is questionable
\end{itemize}

As SMEs rank the importance of each technology, they also provide an assessment of their confidence level for each score. This aids in explicitly defining uncertainty in the evaluation process and prevents false precision. 

\subsection*{Evidence Standards}
All major claims need to be backed up by evidence standards, which stand in place of validation metrics when requirements are not defined. Standards can include mission analogs; operational constraints; after-action reports; documented lessons learned; defined tactics, techniques, and procedures (TTPs) or standard operating procedures; intelligence reporting; specifications or engineering documents. SMEs must provide their evidence standards to tie documentation to any potential design decision so that tribal knowledge is not developed without documentation to support an assumption that was made early on in the process.

\subsection*{Technology Evaluations}
Step 3 of this new approach evaluates the utility of infusing bleeding-edge technologies into a mission solution and establishes a technology utility score. Following the determination of the `how' of a mission, the specific technologies that enable mission success must be identified. All technologies should be considered at this stage, regardless of maturity or development cost. For each critical dimension previously explored in Step 2, as many technologies as possible should be identified that would contribute to progress in that discipline.

Using the standard Technology Readiness Level (TRL) ranking system developed by NASA, technologies are sorted by maturation \cite{noauthor_nasa_nodate}. For this research, technologies in Levels 1--3 are considered ``exploration'' projects, 4--6 ``development'' projects, and 7--9 ``operations'' projects. We will assume that technologies must reach a TRL of 6 --- the final level of the development stage --- by the completion of a program's preliminary design review for it to be feasibly integrated onto a new platform. While technology discipline experts should be included in conversations about the future use cases of their capabilities, it is the mission engineering experts who should conduct utility assessments due to their broader knowledge of the mission context. This comprehensive assessment will use a modification of Girard's Multi-Attribute Utility Theory (MAUT) to quantify the utility of a system based on a series of probabilistic attributes \cite{p_e_girard_modeling_nodate}. The modified assessment conducted by ME SMEs should, at a minimum, include considerations for:
\begin{itemize}
    \item Affordability -- what is the predicted cost to mature the technology to a TRL of 6?
    \item Effectiveness -- how much does the mission feasibility improve with the inclusion of this technology?
    \item Flexibility -- what level of interoperability does the technology have with other systems?
    \item Suitability -- how easily can the technology be used in an operational context?
    \item Multiplicity -- how will utilizing multiple instances of this system increase its effectiveness?
\end{itemize}

\subsection*{Step 3.1: Define the Mission Utility Model}
Using the narrative definition of success previously developed and the top-ranked critical dimensions, this step provides a mathematically consistent way to transform heterogeneous measures into a single value metric. In a model-based implementation, the mission utility model is represented using SysML value properties, constraint blocks, and parametric relationships linked back to the UAF mission architecture. The critical dimension weights calculated in Step 2.4, performance variables, and utility curves are all treated as model elements, so changes in any assumptions or performance predictions can be easily propagated through the calculations. From Girard's MAUT framework, all attributes should be non-redundant, operational, and as independent as is feasible. The top-level critical dimensions and their subsequent weights $w_i$ are abstract and non-operational, so we define mission effectiveness attributes $j$ that are decision-relevant and operational for each critical dimension that map to a utility $u_j (x_j) \in [0,1]$ where $x_j$ is the performance variable. For the utility model, each selected critical dimension is represented by at least one mission effectiveness attribute indexed by $j$. The weight $w_j$ is the mission priority weight assigned to that operationalized attribute. At this stage of the analysis, each utility should be a piecewise curve to account for preferential meaning and make tradeoffs explicitly defined.

For the CAS example, for an attribute related to agility, $x =$ \textit{minutes from  call to first effects on target}, where a smaller value is better performance, a utility curve may be defined as: 
\begin{align} \label{eq:utility-curve}
\raisebox{-36pt}{\makebox[0pt]{\smash{\input{utility-example}}\hspace{2in}}}
u(x) =\begin{cases}
1, &  x \leq 5\\
1-0.3 \frac{x-5}{5},  & 5 < x < 10\\
0.7-0.4 \frac{x-10}{10},  &  10 < x < 20\\  
0.3-0.3 \frac{x-20}{10},  &  20 < x < 30\\  
0,  &  x \geq  30
\end{cases}
\end{align}

The baseline mission utility is then calculated to be used as a reference against technology infusions in subsequent steps. This anchors the assessment with currently feasible performance and ensures that technology impacts can be measured as deltas from a stable reference point. It also allows for future scenario analysis where the baseline can be used as a reference architecture and the assumptions can be varied for trade studies. For $m$ total number of operational attributes, the baseline mission utility $U_{base}$ is described by:
\begin{align} \label{eq:baseline-utility}
    U_{base} = \sum_{j=1}^m w_j u_j (x_j)
\end{align}

This step is complete when all attribute utility curves are defined, as well as the baseline mission utility.

\subsection*{Step 3.2: Identify Discriminating Technologies}
Since mission complexity is primarily driven by technologies whose performance directly affects both mission utility and feasibility, we must next define the technologies that provide mission-critical impacts. A valid technology for this step should be able to meaningfully change the baseline mission utility via the performance variable $x$ improvement of at least one attribute $u$. Alternatively, a technology could also prevent a redline failure or dependency. For each technology $k$, the analyst defines the current TRL, mission functions from Step 2 of the MFA it enables, the attributes it affects, if it is a redline or single-point dependency, if it requires the integration of other technologies, and if alternatives to this technology exist. This information all informs how critical a technology is to the mission and will help quantify its impact in later steps. This step is complete when each relevant technology is captured as a technology element, resource block, or system block with associated TRL, mission functions, affected utility attributes, dependencies, alternatives, evidence standards, and justification.

\subsection*{Step 3.3: Estimate Technology Mission Utility Leverage}
Next, to ensure traceability through the process, the technologies are linked to utility through their change to attribute performance. For $n$ number of technologies $k$, we quantify the new performance of each affected attribute $x_{j,k}$. We then calculate an attribute-level utility delta:
\begin{align} \label{eq:utility-delta}
   \Delta u_j = u_j (x_{j,k}) - u_j (x_{j,base})
\end{align}

The technology utility leverage $L_k$, or the amount of value the technology buys across the entire mission utility, is calculated as the sum of utility deltas applied to each attribute it impacts, $A_k$:  
\begin{align}\label{eq:utility-leverage}
   L_k = \sum_{j \in A_k} w_j \Delta u_j
\end{align}

This step is essential for documenting which technologies are most impactful to the overall mission. An updated mission utility is then the sum of baseline utility and the new technology utility leverage. This creates a traceable relationship from each technology to the performance variable it changes, the resulting utility delta, and the updated mission utility. This step is complete when both a utility leverage and an updated mission utility are generated.

\subsection*{Step 3.4: Quantify Technology Complexity}
 
 A mission complexity factor should increase when high mission utility depends on technologies with low maturity, weak evidence, or low confidence in their claimed effects. To convert TRL to a maturity deficit $M$, where a TRL 1 technology receives a score of 1, and a TRL 9 a score of 0, we use the following:
\begin{align} \label{eq:maturity-deficit}
   M = \frac{9-TRL}{8}
\end{align}

We then combine utility leverage with TRL and expert-assessed evidence standards and confidence to compute a technology complexity (TC) value, which shows the technology's contribution to mission complexity as a value-weighted risk term:
\begin{align} \label{eq:technology-complexity}
   TC_k = \mathrm{max}(0,L_k) \left[ \alpha M_k + \beta (1-C_k) + \gamma (1-E_k) + \delta I_k \right]
\end{align}

\noindent where $C$ is the confidence in the estimated utility impact between [0,1] (0 is no confidence, and 1 is certainty), $E$ is the quality of the evidence that supports the utility impact from [0,1] (0 being analysis-only to 1 being operational demonstration), and $I$ is the integration or coupling complexity from [0,1] (0 being no integration challenges or dependency on other technologies and 1 being a fully dependent relationship or extreme integration challenges).

All coefficients, $\alpha$, $\beta$, $\gamma$, and $\delta$ are tunable weights. The $\max$ term puts a higher value on technologies that are necessary for the mission. The overall equation demonstrates that mission complexity will grow when mission success is dependent on low-maturity, weakly supported, low-confidence, or tightly coupled technologies. In a SysML implementation, TRL, maturity deficit, confidence, evidence quality, integration coupling, and utility leverage are all modeled as value properties connected to a technology complexity constraint block. This step is complete when the technology complexity factor, $TC_k$, is calculated for each candidate technology.

\subsection*{Step 3.5: Calculate Mission Complexity Factor}
Following the quantification of each technology's utility to the mission objectives, Step 3.5 combines the assessments from the two previous steps to develop an overall Mission Complexity Factor (MCF):
\begin{align} \label{eq:mcf}
   MCF = 100 \left[ \lambda D + (1-\lambda) \frac{\sum_{k=1} ^n TC_k}{\sum_{k=1} ^n \mathrm{max}(0,L_k) + \eta} \right]
\end{align}

\noindent where $D$ is the mission inherent difficulty score from Step 2.3, $\lambda \in [0,1]$ is a tunable parameter controlling the relative importance of mission difficulty and technology complexity, and $\eta$ is a small positive constant to prevent division by zero.
For a first-order analysis, a morphological matrix method should be used to construct combinations of candidate mission architecture alternatives and groupings of compatible technologies that enable mission success. This morphological matrix method closely mirrors a similar process conducted in early-stage aircraft design to identify potential aircraft configurations. MCF is then calculated for each architecture alternative/technology group combination using the MFA and TC scores previously defined. For each mission, the ideal architecture alternative is the one with the lowest MCF, indicating it is the least complex solution that will complete the task. To guide the requirements generation process, the architecture alternative for the most important mission is used as the primary architecture for Tier 1 requirements.

\section{Model-Based Requirement Derivation Methodology (Step 4)}
Finally, Step 4 from Figure~\ref{fig:activity} derives model-based Tier 1 requirement candidates from the prioritized mission set and the selected architecture alternative using the previous feasibility and complexity assessments. During the conceptual design phase, Tier 1 requirements should trace directly to the mission objective through platform-agnostic mission functions, constraints, mission effectiveness attributes, utility curves, and technologies. With this method, UAF provides the mission architecture framework for organizing threats, environmentals, CONOPS, interfaces, operational performers, and mission threads. SysML provides the modeling implementation to represent those elements and connect them to requirement, structural, behavioral, and parametric artifacts. This ensures that both physical and functional requirements are traceable to the high-level mission objective.

Each derived requirement should have explicit traceability to the model elements that justify it. Functional requirements are derived from mission functions and activities, while performance requirements are derived from effectiveness attributes and utility curves. Any interface and interoperability requirements are derived from technology coupling, resource dependencies, and operational exchanges. In a SysML implementation, these relationships should be represented using <<deriveReqt>>, <<satisfy>>, <<trace>>, and <<verify>> relationships. 
 
Per DoD instruction, Tier 1 requirements become the driving KPPs for the program \cite{lescher_top_2021}. In this new approach, the least complex architecture alternative is used to develop the initial iteration of Tier 1 requirements and associated KPPs. This sequence of analyzing mission difficulty, technology feasibility, and overall mission complexity is critical; different architecture alternatives can produce vastly different Tier 1 requirement sets using this method, and therefore downstream architecture effects. Without the previous screening steps, there is potential to add unnecessary technical risk into the baseline design that will not be realized until it is costly to pivot. For example, a mission could use stealth or overwhelming force to accomplish the same task, but the requirements for both architecture alternatives would be very different. 

For Tier 2 requirements, the MAUT analysis and utility curves should be used to define threshold and objective values for supporting performance variables based on required mission effectiveness. The technology mission utility leverage values developed in the previous steps provide reasonable bounds of capability increase that can be allocated above the baseline. These supporting values can be captured as lower-level SysML requirement elements traced up to Tier 1 requirement candidates. The definition of requirements below Tier 2 is out of the scope of this research.

\section{Notional Example: Close Air Support}

This section demonstrates the proposed framework using a notional close air support (CAS) mission. The example is intended to show how an ambiguous mission intent can be systematically decomposed. All numbers are notional and included only to illustrate the mechanics of the approach; they do not represent validated data or technology estimates. 

\subsection*{Step 1: Mission Definition}
The first step is to establish the mission objective and context required to define success. Although the objective is straightforward at the highest level, the mission can differ significantly depending on the mission context. With this in mind, select contextual factors are also defined below to develop a more robust mission basis.

\begin{itemize}

\item \textbf{Mission objective:}~provide close air support to ground troops

\item \textbf{Ground troop mission:}~delay, breach, urban clearance, convoy movement

\item \textbf{Operational performers:}~airborne platform, command and control (C2) node, Joint Terminal Attack Controller (JTAC), ground unit

\item \textbf{Threat context:}~adversary air defenses, jamming, airspace control, small arms capabilities, expected threat density and locations

\item \textbf{Deconfliction considerations:}~medical evacuation routes, ally weapons firing zones, other CAS vehicles, autonomous vehicles, how deconfliction is communicated

\item \textbf{Abort criteria:}~loss of communication, extreme weather, weapons malfunction

\item \textbf{Target prioritization:}~armored vehicles, infantry, compounds, military infrastructure

\item \textbf{Rules of engagement:}~positive identification requirements, collateral thresholds, civilian presence, approval authorities

\end{itemize}

\noindent Together, these elements build the initial mission context for a CAS scenario, providing a narrative and architectural foundation for the feasibility assessment that follows.

\subsection*{Step 2: Mission Feasibility Assessment}
The second step translates mission context into success criteria, mission functions, constraint levels, and a normalized inherent difficulty score. It intentionally evaluates a mission at a platform-agnostic level prior to leveraging technologies or platform solutions to best understand what the mission requires. 

\subsubsection*{Step 2.1: Narratively Define Success}
First, the definition of a successful mission is developed to provide a qualitative basis for evaluating whether an architecture can complete the intended mission. 

\textbf{Success statement:} Support is successful if it arrives within the ground commander's decision window, provides the intended munitions and effects, and does not create unacceptable civilian or ground troop risk

\textbf{Additional success definitions: }
\begin{itemize}
    \item Target validation meets established confidence standards
    \item Mission abort remains available throughout the duration of the attack window
    \item Deconfliction with other assets can be maintained at nominal complexity levels
\end{itemize}

This success statement and associated definitions helps to decompose the CAS objective into initial mission criteria. In the next steps, qualitative concepts in the definitions will be further refined into mission functions, constraints, and utility attributes. 

\subsubsection*{Step 2.2: Define Platform-Agnostic Mission Functions and Constraints}
Next, the mission is decomposed into the functions required to physically execute the objective. These are still platform-agnostic and do not allude to specific aircraft, weapons, sensors, or other systems. The functions are then paired with a constraint that could hinder execution. The resulting mission function and constraints, shown in Table~\ref{tab:step 2.2}, provide a basis for rating the inherent mission difficulty in the next step.
These functions represent the major execution steps to complete the CAS mission, and their associated constraints identify where the mission could become more difficult. 

\begin{table}[hbt!]
    \centering
\caption{Close Air Support Mission Functions and Constraints}
\label{tab:step 2.2}
    \begin{tabular}{|c|c|c|c|}\hline
        \textbf{ Mission Function}&  \textbf{Function Description}&  \textbf{Constraint}& \textbf{Constraint Level}\\\hline
         Access&  Enter contested mission area&  Airspace permissiveness& High\\\hline
         Coordinate&  Talk to JTAC + C2 + other assets&  Human dependency& High\\\hline
         Sense&  Collect intelligence&  Environment severity& Low\\\hline
         Decide&  Prioritize and authorize engagements&  Information quality& Medium\\\hline
         Deliver&  Apply effects&  Collateral damage& High\\\hline
         Recover&  Return to base&  Failure tolerance& Medium\\ \hline
    \end{tabular}
    
\end{table}

\subsubsection*{Step 2.3: Rate Mission Inherent Difficulty}
The next step is to score the mission's inherent difficulty level using the constraint levels from Step 2.2 and some additional mission-level factors. The score represents the difficulty prior to any technology infusions or mitigation solutions. A high value for difficulty indicates the mission is intrinsically challenging due to factors including external environmentals, coordination complexity, uncertainty, or the severity of failure consequences. 

\begin{table}[hbt!]
\centering
\caption{Constraint Difficulty Score}
\label{tab:step 2.3 constraint}
\begin{tabular}{| l  |l  |l |}
\hline
\textbf{Constraint} & \textbf{Level} & \textbf{Score} \\
\hline
Airspace permissiveness & High& 5 \\
\hline
Human dependency & High& 5 \\
\hline
Environment severity & Low& 1 \\
\hline
Information quality & Medium& 3 \\
\hline
Collateral damage & High& 5 \\
\hline
Failure tolerance & Medium& 3 \\
\hline
 \multicolumn{2}{|r|}{\textbf{Total Score}}&22/30\\ \hline

\end{tabular}
    
\end{table}

The constraint difficulty score in Table~\ref{tab:step 2.3 constraint} shows that airspace permissiveness, human dependencies, and collateral damage have a large impact on the overall success of the mission. All these of these factors should have appropriate mitigation strategies to minimize their impact. 

\begin{table}[hbt!]
\centering
\caption{Mission Factors Difficulty Score}
\label{tab:step 2.3 factors}
\begin{tabular}{| l  |l  |l |}
\hline
\textbf{Mission Level Factor} & \textbf{Rationale} & \textbf{Score} \\
\hline
Operational complexity & Rapidly changing ground picture, multiple handoffs & 3 \\
\hline
Uncertainty/ambiguity & Target identification, environmentals & 4 \\
\hline
Consequence of failure & Loss of friendly troops or civilians & 5 \\
\hline
 \multicolumn{2}{|r|}{\textbf{Total Score}}&12/15\\ \hline

\end{tabular}

\end{table}

The mission-level factors in Table~\ref{tab:step 2.3 factors} represent the broader contributors that are not fully captured by individual constraints. Operational complexity covers the need for multiple handoffs in communication and authority alongside a constantly evolving ground picture; uncertainty reflects any unknowns or low-confidence information; the consequence of failure refers to the severity of potential losses of life if support cannot be provided or weapon effects veer off-target.

Together, the total mission's inherent difficulty level becomes: 
\begin{align*}
   D = \frac{S_{constraints}+S_{mission factors}}{S_{max}} = \frac{22+12}{30+15} = 0.755
\end{align*}

\noindent So, for this mission, the notional, normalized inherent difficulty level $D$ is $0.755$. This value indicates a relatively difficult mission and provides one of two inputs for the mission complexity factor calculation in Step 3.5.

\subsubsection*{Step 2.4: Rank Critical Dimensions using MaxDiff}
The next step is to define which critical dimensions are most important to successfully executing the mission objective. For a CAS mission, five potential critical dimensions are: survivability, agility, lethality, affordability, and maneuverability. Because these dimensions are not assumed to be of equal importance, a MaxDiff survey of SMEs is used to estimate relative mission weights. 

Using notional results of a MaxDiff survey of 5 SMEs, the weights for the critical dimensions can be calculated using Equation~(\ref{eq:wi}).
Table~\ref{tab:step 2.4} shows that survivability was ranked as the most important dimension, with lethality second most important. The results are consistent with the operational picture for a CAS mission, where success depends heavily on the platform surviving in the threat environment to be able to deliver effects. These weights will be used in the subsequent step to ensure that technology comparisons are scaled to reflect mission priorities.

\begin{table}[hbt!]
    \centering
\caption{Weighted Importance of CAS Critical Dimensions}
\label{tab:step 2.4}
    \begin{tabular}{|c|c|c|c|}\hline
        \textbf{ Critical Dimension, $i$}&  \textbf{Latent Utilities, $v_i$}&  \textbf{Exponential, exp($v_1$)}& \textbf{Weight, $w_i$}\\\hline
         Survivability&  1.2&  3.3201& 0.329\\\hline
         Agility&  0.6&  1.8221& 0.181\\\hline
         Lethality&  0.8&  2.2255& 0.221\\\hline
         Affordability&  0.2&  1.2214& 0.121\\\hline
         Maneuverability&  0.4&  1.4918& 0.148\\ \hline
    \end{tabular}

\end{table}

\subsection*{Step 3: Mission Complexity Factor}
Step 3 of the approach evaluates how potential new technologies can affect the mission as well as the level of complexity involved in their integration. The final result is a mission complexity factor that encompasses both the inherent difficulty of the mission and the complexity of the technologies needed to improve mission utility. 

\subsubsection*{Step 3.1: Define the Mission Utility Model}
A mission utility model is used to translate each weighted critical dimension into measurable attributes with performance values and a corresponding utility value. This provides a baseline utility to compare against when technology impacts are evaluated. The curve also allows for trade studies to be incorporated into an analysis structure to understand how changing the performance will affect overall mission utility.
Using the utility curve $u(x)$ shown in Step 3.1 in Equation~(\ref{eq:utility-curve}) for the mission effectiveness attribute related to the agility of 'time to effects', we can compute the baseline mission utility.
% , the following utility curve can be defined: 
% \begin{align}
%     u(x) =\begin{cases}
% 1, &  x \leq 5\\
% 1-0.3 \frac{x-5}{5},  & 5 < x < 10\\
% 0.7-0.4 \frac{x-10}{10},  &  10 < x < 20\\  
% 0.3-0.3 \frac{x-20}{10},  &  20 < x < 30\\  
% 0,  &  x \geq  30
% \end{cases}
% \end{align}
Using the critical dimensions identified in Step 2.4 and performance values for each effectiveness attribute shown in Table~\ref{tab:step3.1}, utility curves are created using piecewise curves. The table operationalizes each critical dimension to at least one effectiveness attribute, although multiple attributes could be associated with the same critical dimension if desired. Each attribute inherits the weight associated with its parent critical dimension. The utility value for each attribute is then used to calculate a baseline mission utility.
\begin{table}[hbt!]
\centering
\caption{Notional Baseline Mission Utility Information}
\label{tab:step3.1}
\begin{tabular}{| >{\raggedright}p{1.4cm} | >{\raggedright}p{2.4cm} | l  | >{\raggedright}p{1.2cm}  | >{\raggedright}p{2cm} | >{\raggedright\arraybackslash}p{1.1cm} |}
\hline
 \textbf{Attribute ID} & \textbf{Critical Dimension} & \textbf{Effectiveness Attribute} & \textbf{Weight, $w_j$}& \textbf{Performance, $x_j$}& \textbf{Utility, $u_j$}\\
\hline
 $A_1$&Survivability & Probability to survive [\%]& 0.329 & 95 & 0.55 \\
\hline
 $A_2$&Agility & Time to effects [min] & 0.181 & 15 & 0.50 \\
\hline
 $A_3$&Lethality & Probability of effect success [\%]& 0.221 & 70 & 0.60 \\
\hline
 $A_4$&Affordability & Cost per sortie [low-high] & 0.121 & 0.5 & 0.50 \\
\hline
 $A_5$&Maneuverability & Minimum turn radius [nmi] & 0.148 & 0.75 & 0.70 \\
\hline

\end{tabular}

\end{table}

The baseline mission utility is then calculated using Equation~(\ref{eq:baseline-utility}): 
\begin{align*}
   U_{base} = (0.329*0.55)+(0.181*0.5)+(0.221*0.6)+(0.121*0.6)+(0.148*0.7) = 0.568
\end{align*}

This baseline value is the reference point for evaluating the benefits of any new technologies in an architecture alternative. Candidate technologies are judged both by how they improve a single performance variable and also by how much value they provide relative to the baseline mission utility.

\subsubsection*{Step 3.2: Identify Discriminating Technologies}
After the baseline utility is established, the next step is to identify any technologies that could improve the mission's utility or reduce a constraint or dependency. The technologies selected for this example are not exhaustive or revolutionary; they are only used to demonstrate the process of how a technology can be linked to the greater mission context. 

Table~\ref{tab:step 3.2} links four candidate technologies to the mission functions they support and the corresponding effectiveness attributes it affects. These links establish traceability required to calculate technology utility leverages in the subsequent step. 

\begin{table}[hbt!]
\centering
\caption{Discriminating Technologies and Impact Information}
\label{tab:step 3.2}
\begin{tabular}{|l |l |l | >{\raggedright}p{1.6cm} | l|}\hline
\textbf{Tech ID} & \textbf{Technology $k$}& \textbf{Mission Functions} &\textbf{Attributes Impacted}& \textbf{TRL} \\\hline
$k_1$ & Jamming-resilient C2/targeting & Coordinate / Decide  &$A_1$, $A_3$& 5 \\\hline
$k_2$ & Low-collateral precision effects & Deliver  &$A_3$, $A_4$& 7 \\\hline
$k_3$ & Automatic target recognition-assisted target validation & Sense / Decide  &$A_2$, $A_3$& 6 \\\hline
$k_4$ & Autonomous deconfliction & Coordinate / Recover  &$A_1$, $A_3$& 4 \\ \hline

\end{tabular}

\end{table}

\subsubsection*{Step 3.3: Estimate Technology Mission Utility Leverage}
For each of the four technologies identified in Step 3.2, the utility deltas are calculated using Equation~(\ref{eq:utility-delta}) to estimate how much the technology changes mission utility. The accompanying technology leverages are calculated using Equation~(\ref{eq:utility-leverage}) to show the weighted utility gained by improving all attributes related to a given technology. Technology $k_1$ is shown below as an example. 

\textbf{Utility Delta:} We assume that the technology improves the probability to survive from 95\% to 97\% and the probability of effect success from 70\% to 80\%. This results in the following utility deltas, calculated using Equation~(\ref{eq:utility-delta}): 
\begin{align*}
\Delta u_1 = 0.67 - 0.55 = 0.12 \\
\Delta u_3 = 0.75 - 0.60 = 0.25
\end{align*}

\textbf{Technology Leverage:} The leverage for technology $k_1$, which proves jamming-resilient command and control and targeting, is calculated using Equation~(\ref{eq:utility-leverage}): 
\begin{align*}
   L_1 = (0.329*0.12) + (0.221*0.25) = 0.095
\end{align*}

The resulting technology leverages for all four technologies are shown in Table~\ref{tab:step 3.3} below.
In this example, autonomous deconfliction provides the largest impact on mission utility, shown by the largest leverage value. Because they do not have considerations for maturity, integration complexity, confidence of success, or quality of evidence, these values alone should not be used to determine a preferred technology set for an architecture.

\begin{table}[hbt!]
\centering
\caption{Summary of Technology Leverages}
\label{tab:step 3.3}
\begin{tabular}{| l | l | l | l |}
\hline
\textbf{Tech ID} & \textbf{Attributes Impacted, $A_k$} & \textbf{Utility Deltas, $\Delta u_j$}  & \textbf{Technology Leverage, $L_k$} \\
\hline
$k_1$ & $A_1$, $A_3$ & 0.12, 0.25 & 0.095 \\
\hline
$k_2$ & $A_2$, $A_4$ & 0.08, 0.15 & 0.033 \\
\hline
$k_3$ & $A_2$, $A_3$ & 0.16, 0.04 & 0.038 \\
\hline
$k_4$ & $A_1$, $A_3$ & 0.2, 0.18 & 0.11 \\
\hline

\end{tabular}

\end{table}

\subsubsection*{Step 3.4: Quantify Technology Complexity}
While the technology leverage quantifies the benefit of a technology, it does not take the risk associated with integrating the technology into account. Step 3.4 calculates the overall technology complexity value by incorporating confidence that the technology can provide the benefit it claims, the quality of evidence that supports the technology development, the current maturity level, and integration difficulty. This ensures a design decision based not only on the potential mission benefits but also the risk of including the technology as a core pillar of an architecture alternative. 
Using Equation~(\ref{eq:maturity-deficit}), the maturity deficit $M$ for a TRL 5 technology like $k_1$ is 0.5. Assuming a confidence level of 0.60, evidence quality of 0.8, integration complexity of 0.55, and the following tuning weights: 
\begin{align*}
   \alpha = 0.25, \quad \beta = 0.20, \quad \gamma = 0.20, \quad \delta = 0.35
\end{align*}

\noindent The technology complexity $TC_1$ for $k_1$ can be calculated using Equation~(\ref{eq:technology-complexity}): 
\begin{align*}
   TC_1 = \max(0,0.095)[(0.25*0.5)+0.2(1-0.6)+0.2(1-0.80)+(0.35*0.55)] = 0.0416
\end{align*}

The resulting technology complexities for all technologies are shown in Table~\ref{tab:step 3.4} below, where we can see the importance of including complexity factors into the analysis rather than just the mission utility. While autonomous deconfliction has the largest utility leverage, it also has the highest complexity value due to low maturity, confidence, and evidence basis, along with a high integration complexity. This shows that while it provides a large benefit, it is also a high risk to incorporate into an architecture alternative. 

\begin{table}[hbt!]
\centering
\caption{Technology Assumptions and Resulting Complexities}
\label{tab:step 3.4}
\begin{tabular}{|l |l |l |l |l |l |l |l|}\hline
\textbf{Tech ID} & \textbf{TRL} & \textbf{$M_k$} & \textbf{$C_k$} & \textbf{$E_k$} & \textbf{$I_k$} & \textbf{$L_k$} & \textbf{Technology Complexity, $TC_k$} \\\hline
$k_1$ & 5 & 0.500 & 0.60 & 0.80 & 0.55 & 0.095 & 0.0416 \\\hline
$k_2$ & 7 & 0.250 & 0.70 & 0.70 & 0.40 & 0.033 & 0.0106 \\\hline
$k_3$ & 6 & 0.375 & 0.55 & 0.50 & 0.60 & 0.038 & 0.0188 \\\hline
$k_4$ & 4 & 0.625 & 0.45 & 0.40 & 0.80 & 0.11 & 0.0732 \\ \hline

\end{tabular}

\end{table}

\subsubsection*{Step 3.5: Calculate Mission Complexity Factor}
Finally, the mission complexity factor can be calculated using the mission difficulty score from Step 2.3, the leverage values from Step 3.3, and the technology complexity values from Step 3.4. The resulting score describes the overall complexity of the mission for a given architecture alternative. A low MCF indicates a less complex architecture for completing the mission objective, while a higher value indicates a more complex architecture to execute the same mission. 
Assuming a tuning value of $\lambda=0.4$ and a small constant $\eta=1*10^{-6}$, the overall mission complexity factor for this mission architecture alternative can now be calculated using Equation~(\ref{eq:mcf}): 
\begin{align}
   MCF = 100 \left[ (0.4*0.755)+ (1-0.4) \frac{(0.0416+0.0106+0.0188+0.0732)}{(0.095+0.033+0.038+0.11) + 1*10^{-6}} \right] = 61.55
\end{align}

An MCF of 61.55 represents a moderately complex architecture alternative. In practice, multiple architectures should be evaluated using combinations of technologies and assumptions. Each alternative, then, is evaluated using the same approach and its MCF compared to the other architectures. The architecture with the lowest MCF for the highest-priority mission then becomes the baseline for deriving Tier 1 requirements.

\section{Conclusion}
This work proposed a structured approach for deriving mission engineering requirements when customer requirements are incomplete or unavailable. The new method connects early mission definition and requirements engineering practices by decomposing mission context into traceable artifacts. An initial mission feasibility assessment establishes the difficulty of completing the mission without the infusion of new technology and considers the potential impacts of external factors. A MaxDiff weighting method was proposed to identify the most relevant critical dimensions for the mission as evaluated by subject matter experts. Next, a MAUT-inspired utility model was developed to provide a quantitative baseline to compare candidate technology impacts. A mission complexity factor is defined to balance mission difficulty, technology potential, and integration risk. Throughout each step of the proposed approach, UAF/UAFML and SysML artifact mappings are identified. This approach supports traceability from initial mission context and objectives to model-based Tier 1 and Tier 2 requirement definition. The notional example shows how the process can be applied to a relevant defense mission scenario to develop requirements in the absence of customer input. Future research will incorporate this approach into an MBSE environment with all necessary artifacts. 

\bibliography{references_short}

\end{document}

%% file: utility-example.tex
% \scalebox{0.5}{%
\begin{tikzpicture}
\begin{axis}[
    axis lines=middle,
    xlabel={$x$},
    ylabel={$u(x)$},
    ylabel style={yshift=1.3em},
    xmin=0, xmax=35,
    ymin=0, ymax=1.1,
    xtick={0,5,10,20,30},
    ytick={0,0.3,0.7,1},
    grid=both,
    width=1.8in,
    height=1.4in,
    domain=0:35,
    samples=2,
]
% \normalsize
\small

% u(x) = 1, x <= 5
\addplot[blue, thick, domain=0:5] {1};

% u(x) = 1 - 0.3(x-5)/5, 5 < x < 10
\addplot[blue, thick, domain=5:10] {1 - 0.3*(x-5)/5};

% u(x) = 0.7 - 0.4(x-10)/10, 10 < x < 20
\addplot[blue, thick, domain=10:20] {0.7 - 0.4*(x-10)/10};

% u(x) = 0.3 - 0.3(x-20)/10, 20 < x < 30
\addplot[blue, thick, domain=20:30] {0.3 - 0.3*(x-20)/10};

% u(x) = 0, x >= 30
\addplot[blue, thick, domain=30:35] {0};

% Points
\addplot[only marks, mark=*, blue] coordinates {
    (5,1)
    (10,0.7)
    (20,0.3)
    (30,0)
};

\end{axis}
\end{tikzpicture}%
% }